\documentclass[letterpaper,preprintnumbers,prd,twocolumn,nofootinbib,nobibnotes,showpacs]{revtex4}
\usepackage{graphics,epsfig,subfigure}
\usepackage{color}
\usepackage{amsfonts}
\usepackage{mathrsfs}
\usepackage{epsfig}
\usepackage{graphicx}%
\usepackage{dcolumn}
\usepackage{amsmath}
\usepackage{color}
\usepackage{color}

\begin{document}
\renewcommand{\baselinestretch}{1.3}
\newcommand\beq{\begin{equation}}
\newcommand\eeq{\end{equation}}
\newcommand\beqn{\begin{eqnarray}}
\newcommand\eeqn{\end{eqnarray}}
\newcommand\nn{\nonumber}
\newcommand\fc{\frac}
\newcommand\lt{\left}
\newcommand\rt{\right}
\newcommand\pt{\partial}

\allowdisplaybreaks

\title{Black holes of multiple horizons without mass inflation}
\author{Changjun Gao$^{1,2}$\footnote{gaocj@nao.cas.cn}}\author{Toktarbay Saken$^{3,4}$\footnote{toktarbay.saken@kaznu.kz}}

\affiliation{$^1$National Astronomical Observatories, Chinese Academy of Sciences, 20A Datun Road, Beijing 100101, China}
\affiliation{$^2$School of Astronomy and Space Sciences, University of Chinese Academy of Sciences, 19A Yuquan Road, Beijing 100049, China}
 \affiliation{$^3$Al-Farabi Kazakh National University, Almaty 050040, Kazakhstan}
  \affiliation{$^4$Department of Physics, Kazakh National Women’s Teacher Training University, 050000 Almaty, Kazakhstan}

\begin{abstract}
Mass inflation is a phenomenon happened in the vicinity of inner horizon in two-horizon spacetime. It is shown that the mass of initially small perturbations will grow exponentially as they approach the inner horizon. This implies that black hole inner horizon is
unstable to the perturbations. In view of the fact that the mass inflation
is determined by the surface gravity of inner horizon, Carballo-Rubio et al. showed that if one makes the surface gravity of inner horizon vanish, then the exponential growth character of mass inflation is not present. Basing on this conclusion, we look for the black hole solutions of multiple horizons with nonlinear Maxwell field. Then we make the inner horizons to coincide with each other such that the surface gravities of every inner horizon vanish. By this way, black holes of multiple horizons without mass inflation are constructed.  
\end{abstract}






\maketitle


\section{Introduction}

The concept of mass inflation  describes an exponential growth of energy \cite{eric:1990,amos:1991,pat:1995} in the sense that initially small infalling perturbations will grow exponentially as they approach the inner horizon of the two-horizon spacetime.  It is emphasized that the divergence of energy happens in the vicinity of inner horizon. We know that many inner horizons can be in the presence for the multi-horizon black holes.  So the mass inflation phenomenon must occur for  multiple-horizon black holes. Then how to avoid this problem?  Carballo-Rubio et al \cite{rubio:2022} showed that,  if we make attempt to let the surface gravity of inner horizons vanishes, then the exponential growth character of mass inflation instability is not present.  In practice, they constructed specific geometries for regular black holes without  mass inflation and analyze their behavior with the  double null shell method and the modified Ori method, showing that the exponential growth of mass inflation disappears with the vanishing of surface gravity on the inner horizon.  On the other hand, Eichhorn  and Fernandes  \cite{eich:2025} derive regular black hole solutions very recently.  Because the solutions have vanishing surface gravity, they are not susceptible to mass-inflation instability.  

Actually,  in terms of a suitable null coordinate v, the mass inflation is given by \cite{matt:2024}
\begin{eqnarray}
\delta{m}\left(v\right)\propto{\left(v-v_0\right)^{-p}}e^{\left|\kappa_{in}\right|\left(v-v_0\right)}\;,
\end{eqnarray}
where $\kappa_{in}$ is the surface gravity of inner horizon.  It is obvious that the mass inflation
is determined by the inner horizon surface gravity $\kappa_{in}$ . This suggests that if we let $\kappa_{in}$ vanishes, thereby  switching off the mass inflation.  

 How to get the surface gravity of inner horizon vanish? Note that when the two horizons of Reissner-Nordstr$\ddot{o}$m space coincide, the surface gravity of overlapped horizon is exactly zero and the mass inflation is erased. Inspired by this fact, we apply ourselves to the seeking for black holes with multiple horizons. Then we require that  surface gravities for all the inner horizons vanish by coinciding the inner horizons. We shall work in the frame of Einstein nonlinear-Maxwell theory.  
 
 Black holes with many horizons are mainly studied in Refs.~\cite{noji:2017,gao:2018,gao:2021}. Concretely, by using the solution-generating method, Nojiri and Odintsov \cite{noji:2017} present the regular black holes with multi-horizons in modified gravity with nonlinear Maxwell field.   The method of solution-generating denotes that one  presupposes the metric functions initially and then solve for the Maxwell field and the Lagrangian function.  Also  using the solution-generating method, we  obtain the solutions for static and spherically symmetric black holes and cosmos with multiple horizons and multiple singularities in the generalized Einstein-Maxwell theory \cite{gao:2018}.   On the other hand, by using the method of solving the equations of motion straightforwardly,  we find the multi-horizon black hole solutions  basing on the Lagrangian with series of Maxwell term, $F_{\mu\nu}F^{\mu\nu}$  \cite{gao:2021}. In this paper, we shall extend the study to more general case. Specifically, we take the Lagrangian for the Maxwell field as the series of $\sqrt{-F_{\mu\nu}F^{\mu\nu}/2}$ , rather than  $F_{\mu\nu}F^{\mu\nu}$. The former is obviously different from the latter.  One of the benefits from the extension is the emergence of some new terms in the line element.   Those new terms modify the structure of spacetime significantly.  The other solutions for  black holes, black branes, magnetic black holes and magnetic branes with nonlinear Maxwell field are found in Refs.~from \cite{50:1} to \cite{50:25}.

The paper is organized as follows. In Section II, starting from the most general Lagrangian  for the nonlinear Maxwell field, we derive the multiple-horizon black holes by using the series-expansion method. In Section III,  using the solution-generating method, we obtain the Lagrangian for the three-horizon black holes with and without the cosmological term. In Section IV, we explore the thermodynamics of 7-horizon black holes. In practice, the first law of thermodynamics and the Smarr relation are found. In Section V and Section VI, we consider the motion of test particles and test field in the background of multiple-horizon spacetime. In the last Section, we present the  solutions without  mass inflation. Throughout the  paper, we adopt the
system of units in which $G=c=\hbar=1$ and the metric signature
$(-, +, +, +)$.
\section{exact solutions}
In this section, we seek for the exact solutions for multi-horizon black holes in the Einstein nonlinear-Maxwell theory. The theory we consider has the following four dimensional action  
\begin{equation}
S=\int d^4x\sqrt{-g}\left(-\frac{1}{4}R-\frac{1}{4}F^{2}+\sum_{n=3}^{\infty}a_n\chi^n\right)\;,
\end{equation}
where 
\begin{eqnarray}
F^2&=&F^{\mu\nu}F_{\mu\nu}\;,\ \ \ \ \ \ \ \ \ \   \chi=\sqrt{-\frac{1}{2}F^2}\;,\ \ \ \ \ \nonumber\\ F_{\mu\nu}&=&\nabla_{\mu}A_{\nu}-\nabla_{\nu}A_{\mu}\;,
\end{eqnarray}
with $A_{\mu}$ the four-vector Maxwell field.  $R$  is the Ricci scalar. The action has the property of gauge invariant, 
\begin{equation}
A_{\mu}\xrightarrow{}A_{\mu}+\nabla_{\mu}{\psi}\;,
\end{equation}
with $\psi$ an arbitrary scalar field. The Lagrangian $L$ here for the nonlinear Maxwell field is different from that in Ref.~\cite{gao:2021} where the Lagrangian is the series of Maxwell term, $F_{\mu\nu}F^{\mu\nu}$.  The corresponding Lagrangian there is equivalent to 

\begin{equation}
L=-\frac{1}{4}F^{2}+\sum_{n=3}^{\infty}a_n\chi^n+\lambda\left(2\chi^2+F^2\right)\;,
\end{equation}
where $\lambda$ plays the role of auxiliary scalar field.

It has been shown \cite{gao:2018} that the four dimensional static spherically symmetric spacetime sourced by the nonlinear Maxwell field can always be written as  
\begin{equation}\label{metric-1}
ds^2=-U\left(r\right)dt^2+{U\left(r\right)^{-1}}dr^2+r^2d\Omega_2^2\;.
\end{equation}
Namely, there is only one freedom of  $U(r)$ to be determined. Since the spacetime is static and spherically symmetric, the four-vector can be assumed as 
\begin{equation}\label{four-vector}
\left[A_{\mu}\right]=\left[\phi\left(r\right)\;,0\;,0\;,0\right]\;.
\end{equation}
Here $\phi$ corresponds to the Coulomb potential in the Maxwell electrostatics. Let 
\begin{equation}
\phi=\int{K\left(r\right)}dr\;,
\end{equation}
we obtain the following independent equations from the Einstein equations 
\begin{equation}\label{eq:K1}
r^2K+\sum_{n=3}^{\infty}nr^2a_nK^{n-1}+Q=0\;,
\end{equation}
\begin{equation}\label{eq:U1}
\frac{1}{2}-\frac{1}{2}U-\frac{1}{2}rU^{'}-\frac{1}{2}K^2r^2-\sum_{n=3}^{\infty}\left(n-1\right)r^2a_nK^n=0\;.
\end{equation}
$K$ has the physical meaning of minus electric field intensity in electrostatics. $Q$ is an integration constant which has the physical meaning of electric charge for the whole spacetime.   
We focus on the series solution
\begin{eqnarray} 
K&=& \frac{c_2}{r^2}+\frac{c_3}{r^3}+\frac{c_4}{r^4}+\frac{c_5}{r^5}+\cdot\cdot\cdot\;,\nonumber\\
U&=& d_0+\frac{d_1}{r}+\frac{d_2}{r^2}+\frac{d_3}{r^3}+\frac{d_4}{r^4}+\frac{d_5}{r^5}+\cdot\cdot\cdot\;,
\end{eqnarray}
with $c_i$ and $d_i$ the constants to be determined. 
Substitute them into the equations of motion and let each term of power series (from low to high) for $r$ vanishes. We obtain 
\begin{eqnarray}\label{U}
K&=&-\frac{Q}{r^2}-\frac{3a_3Q^2}{r^4}-\frac{2Q^3\left(9a_3^2-2a_4\right)}{r^6}\nonumber\\&&-\frac{5Q^4\left(27a_3^3-12a_4a_3+a_5\right)}{r^8}\nonumber\\&&-\frac{6Q^5\left(-a_6+15a_5a_3+189a_3^4-126a_4a_3^2+8a_4^2\right)}{r^{10}}\nonumber\\&&-\frac{7Q^6}{r^{12}}\left(a_7-1296a_3^3a_4+192a_3a_4^2-20a_5a_4\right.\nonumber\\ &&\left.-18a_3a_6+180a_3^2a_5+1458a_3^5\right)+\cdot\cdot\cdot\;,\nonumber\\
U&=&1-\frac{2M}{r}+\frac{Q^2}{r^2}+\frac{{2}a_3Q^3}{3r^4}+\frac{Q^4\left(9a_3^2-2a_4\right)}{5r^6}\nonumber\\&&+\frac{{2}Q^5\left(27a_3^3-12a_4a_3+a_5\right)}{7r^8}\nonumber\\&&+\frac{2Q^6\left(-a_6+15a_5a_3+189a_3^4-126a_4a_3^2+8a_4^2\right)}{9r^{10}}\nonumber\\&&+\frac{2Q^7}{11r^{12}}\left(a_7-1296a_3^3a_4+192a_3a_4^2-20a_5a_4\right.\nonumber\\ &&\left.-18a_3a_6+180a_3^2a_5+1458a_3^5\right)+\cdot\cdot\cdot\;.
\end{eqnarray}
Here $M$ and $Q$ are the physical mass and electric charge, respectively. Note that we have used the following facts, $c_2=-Q$, $c_3=0$, $c_4=-3a_3Q^2$, etc. Comparing the two expressions of $K$ and $U$ carefully, we find that the ratio of every two terms with the same power of $r$ is proportional to the charge, $Q$. In other words, every two terms have the common constant factors.  More specifically, we observe that $a_3$ appears in the $r^4$ terms,  $a_4$ (and $a_3$) appears in the $r^6$ terms,  $a_5$ (and $a_4$, $a_3$) appears in the $r^8$ terms,  $a_6$ (and $a_5$, $a_4$, $a_3$) appears in the $r^{10}$ terms, and so on. This tells us we can remove high-order terms of $r$ to any desired order by appropriately choosing the parameters. For example, if we let 
\begin{eqnarray} 
a_4&=&\frac{9}{2}a_3^2\;,\ \ \ \  \Leftarrow{}r^6\ \ \rm{term}\;, \nonumber\\ 
a_5&=&-27a_3^3+12a_4a_3\;,\ \ \ \  \Leftarrow{}r^8\ \ \rm{term}\;, \nonumber\\ 
a_6&=&15a_5a_3+189a_3^4-126a_4a_3^2+8a_4^2\;,\ \  \Leftarrow{}r^{10}\ \rm{term}\;,\nonumber\\ 
\cdot\cdot\cdot&=&\cdot\cdot\cdot\cdot\cdot\cdot\;, 
\end{eqnarray}
we would get  
\begin{eqnarray}\label{eq-K}
K&=&-\frac{Q}{r^2}-\frac{3a_3Q^2}{r^4}\;,
\end{eqnarray}
\begin{eqnarray}\label{eq-U}
U&=&1-\frac{2M}{r}+\frac{Q^2}{r^2}+\frac{{2}a_3Q^3}{3r^4}\;.
\end{eqnarray}

We will see that this spacetime describes a three-horizon black holes. As another example, if we let 
\begin{eqnarray} 
a_8&=&\mathscr{F}_8\left(a_7,a_6,a_5,a_4,a_3\right)\;,\ \ \ \  \Leftarrow{}r^{14}\ \ \rm{term}\;, \nonumber\\ 
a_9&=&\mathscr{F}_9\left(a_8,a_7,a_6,a_5,a_4,a_3\right)\;,\ \ \ \  \Leftarrow{}r^{16}\ \ \rm{term}\;, \nonumber\\ 
\cdot\cdot\cdot&=&\cdot\cdot\cdot\cdot\cdot\cdot\;, 
\end{eqnarray}
we would get  
\begin{eqnarray}\label{U7}
K&=&-\frac{Q}{r^2}-\frac{3a_3Q^2}{r^4}-\frac{2Q^3\left(9a_3^2-2a_4\right)}{r^6}\nonumber\\&&-\frac{5Q^4\left(27a_3^3-12a_4a_3+a_5\right)}{r^8}\nonumber\\&&-\frac{6Q^5\left(-a_6+15a_5a_3+189a_3^4-126a_4a_3^2+8a_4^2\right)}{r^{10}}\nonumber\\&&-\frac{7Q^6}{r^{12}}\left(a_7-1296a_3^3a_4+192a_3a_4^2-20a_5a_4\right.\nonumber\\ &&\left.-18a_3a_6+180a_3^2a_5+1458a_3^5\right)\;,\nonumber\\
U&=&1-\frac{2M}{r}+\frac{Q^2}{r^2}+\frac{{2}a_3Q^3}{3r^4}+\frac{Q^4\left(9a_3^2-2a_4\right)}{5r^6}\nonumber\\&&+\frac{{2}Q^5\left(27a_3^3-12a_4a_3+a_5\right)}{7r^8}\nonumber\\&&+\frac{2Q^6\left(-a_6+15a_5a_3+189a_3^4-126a_4a_3^2+8a_4^2\right)}{9r^{10}}\nonumber\\&&+\frac{2Q^7}{11r^{12}}\left(a_7-1296a_3^3a_4+192a_3a_4^2-20a_5a_4\right.\nonumber\\ &&\left.-18a_3a_6+180a_3^2a_5+1458a_3^5\right)\;.
\end{eqnarray}
Here $\mathscr{F}_i$ denotes the combination of coefficients $a_j$. We will see in the later section it describes in general a 7-horizon black hole. In all, the metric function $U$ in Eq.~(\ref{U})  is the most general solution and can give us the black hole spacetime with arbitrary number of horizons.   

Finally, if we let 
\begin{eqnarray} 
a_3&=&0\;,\ \ \ \  \Leftarrow{}r^4\ \ \rm{term}\;, \nonumber\\ 
a_4&\neq &0\;,\ \ \ \  \Leftarrow{}r^6\ \ \rm{term}\;, \nonumber\\ 
a_5&=&0\;,\ \ \ \  \Leftarrow{}r^8\ \ \rm{term}\;, \nonumber\\ 
a_6&=&8a_4^2\;,\ \  \Leftarrow{}r^{10}\ \rm{term}\;,\nonumber\\ 
\cdot\cdot\cdot&=&\cdot\cdot\cdot\cdot\cdot\cdot\;, 
\end{eqnarray}
we would get  
\begin{eqnarray}\label{EEH}
K&=&-\frac{Q}{r^2}+\frac{4Q^3a_4}{r^6}\;,\nonumber\\
U&=&1-\frac{2M}{r}+\frac{Q^2}{r^2}-\frac{2Q^4a_4}{5r^6}\;.
\end{eqnarray}
To our surprise, the metric is nothing but the  static solution for magnetically charged black holes in Einstein-Euler-Heisenberg gravity \cite{all:2020}.  Therefore, it is  fascinating that  different theories give the same black hole solutions.   Similarly, we can get many other solutions.

\section{three-horizon black hokes with and without cosmological constant}
In this section, we focus on the study of three-horizon black holes with and without the Einstein cosmological constant. Starting from the metric, we shall find the corresponding Lagrangian function for the theory.    

\subsection{The Lagrangian for three-horizon black hokes without cosmological constant }
In this subsection, we construct the Lagrangian of the nonlinear Maxwell field for the three-horizon black hole solution. To this end, we start from the most general action 
\begin{equation}
S=\int d^4x\sqrt{-g}\left[-\frac{1}{4}R+L\left(\chi\right)\right]\;,
\end{equation}
where $L(\chi)$ stands for the arbitrary function of $\chi$. The action gives the Einstein equations
\begin{equation}
G_{\mu\nu}=-\frac{2}{\chi}\frac{dL}{d\chi}F_{\mu\lambda}F^{\ \lambda}_{\nu}-2g_{\mu\nu}L\;,
\end{equation}
and the equation of motion for Maxwell field 
\begin{equation}
\nabla_{\mu}\left(\frac{dL}{{\chi}d{\chi}}F^{\mu\nu}\right)=0\;.
\end{equation}
In the background of static and spherically symmetric spacetime given by Eq.~(\ref{metric-1}), they are reduced to 
\begin{equation}
r^2L_{,\chi}+Q=0\;,
\end{equation}
and 
\begin{equation}
\frac{1}{2}-\frac{1}{2}U-\frac{1}{2}U^{'}r-r^2\chi^2\left(\frac{L}{\chi}\right)_{,\chi}=0\;,
\end{equation}
where $Q$ is an integration constant. Compare the two equations with Eq.~(\ref{eq:K1}) and Eq.~(\ref{eq:U1}), we recognize 
that $K$ is exactly $\chi$ in the background of spacetime Eq.~(\ref{metric-1})
\begin{equation}
\chi=K\;.
\end{equation}
{Now we have three functions, $U$, $\chi$ and $L$, but two independent equations. Thus the equations of motion are not closed. Given anyone of them in advance, then we can solve for the other two. This is exactly the reason why we can re-construct the Lagrangian for the three-horizon black holes}.  
Substituting Eq.~(\ref{eq-K}) and Eq.~(\ref{eq-U}) into above equations of motion, we obtain

\begin{eqnarray}\label{lagforthree}
L&=&\frac{18a_3\chi-1+\sqrt{-\left(12a_3\chi-1\right)^3}}{108a_3^2}\;.
\end{eqnarray}
In the limit of $a_3=0$, it reduces to $L=-\frac{1}{2}F_{\mu\nu}F^{\mu\nu}$ which is the Maxwell case. 
This is precisely the Lagrangian for three-horizon black holes without cosmological constant. We notice that there is only one parameter $a_3$ in the theory. Both the mass $M$ and charge $Q$ are not in the presence. This is a very nice character. We have checked that, starting from this Lagrangian,  we would inevitably get the three-horizon solution by solve the equations of motion straightforwardly.  
When $a_3<0$, the metric function $U$ gives in general three horizons, $\tilde{h}_3>\tilde{h}_2>\tilde{h}_1>0$. As an example, we have 

\begin{eqnarray}
\tilde{h}_{1}=0.08\;,\ \ \ \tilde{h}_2=0.38\;,\ \ \ \tilde{h}_3=1.60\;, 
\end{eqnarray}
when 
\begin{eqnarray}
M=1\;,\ \ \  Q=0.8\;,\ \ \ a_3=-0.1\;. 
\end{eqnarray}

For some specific parameters, two or three horizons would coincide. To show this point, we factorize the $U$ function as follows
\begin{eqnarray}
&&1-\frac{2M}{r}+\frac{Q^2}{r^2}+\frac{\frac{2}{3}a_3Q^3}{r^4}\nonumber\\&&=\left(1+\frac{\tilde{k}_1}{r}\right)\left(1-\frac{\tilde{h}_1}{r}\right)^2\left(1-\frac{s\tilde{h}_1}{r}\right)\;.
\end{eqnarray}
Then we arrive at 
\begin{eqnarray}
\tilde{k}_1&=&\frac{s\tilde{h}_1}{1+2s}\;,  \ \ \ a_3=-\frac{3}{2}s^2\tilde{h}_1\sqrt{\frac{1+2s}{\left(3s^2+2s+1\right)^3}}\;,\ \ \  
\nonumber\\M&=&\frac{\tilde{h}_1\left(1+s\right)^2}{1+2s}\;,\ \ \ \ \  \  Q=\tilde{h}_1\sqrt{\frac{3s^2+2s+1}{1+2s}}\;.\ \ \  
\end{eqnarray}
Let us assume that
\begin{eqnarray}
\tilde{h}_1>0\;,\ \ \ s>0\;.   
\end{eqnarray}
Then we have the following conclusions.  

(1). When 
\begin{eqnarray}\label{eq-oh}
s=1\;,   
\end{eqnarray}
the three horizons $\tilde{h}_3$, $\tilde{h}_2$ and $\tilde{h}_1$ coincide. 

(2). When 
\begin{eqnarray}\label{eq-th}
s>1\;,   
\end{eqnarray}
the two inner horizons $\tilde{h}_2$ and $\tilde{h}_1$  coincide. 

(3). When 
\begin{eqnarray}
0<s<1\;,   
\end{eqnarray}
the event horizon $\tilde{h}_3$ coincide with the inner horizon $\tilde{h}_2$. 
\subsection{Lagrangian for the three-horizon black hokes with cosmological constant}
In this subsection, we present the Lagrangian for the three-horizon black holes with cosmological constant. After we get the Lagrangian Eq.~(\ref{lagforthree}) for three-horizon black holes without cosmological constant, it is straightforward to extend it to the case with the cosmological constant  by simply adding the term for $\Lambda$ 
\begin{eqnarray}\label{lagforthree}
L&=&\frac{18a_3\chi-1+\sqrt{-\left(12a_3\chi-1\right)^3}}{108a_3^2}-\frac{1}{2}\Lambda\;.
\end{eqnarray}
Staring from above Lagrangian, we obtain the black hole solution as follows

\begin{eqnarray}
K&=&-\frac{Q}{r^2}-\frac{3a_3Q^2}{r^4}\;,\nonumber\\
U&=&1-\frac{2M}{r}+\frac{Q^2}{r^2}+\frac{{2}a_3Q^3}{3r^4}-\frac{1}{3}\Lambda{r^2}\;.
\end{eqnarray}
In this case, there are generally four horizons for positive cosmological constant in this spacetime, one cosmic horizon and three black hole horizons.   It is straightforward to obtain the the solutions for Eq.~(\ref{U}) ,  Eq.~(\ref{U7}) and Eq.~(\ref{EEH}) with the cosmological constant by simply adding the term $-\frac{1}{3}\Lambda {r}^2$ into the $U$ function.  

\section{thermodynamics of 7-horizon black hole}
In this section, we shall make a study on the thermodynamics of multiple horizons.  As an example, we shall consider the case of 7-horizon. We find that the positive and negative surface gravities  appear alternately.  It is exactly this character makes us to achieve black holes without mass-inflation. 

The solution for 7-horizon black holes can be written as from Eq.~(\ref{U7})
\begin{eqnarray}
\phi&=&\frac{Q}{r}+\frac{a_3Q^2}{r^3}+\frac{2Q^3\tilde{a}_4}{5r^5}+\frac{5Q^4\tilde{a}_5}{7r^7}\nonumber\\&&+\frac{2Q^5\tilde{a}_6}{3r^{9}}+\frac{7Q^6\tilde{a}_7}{11r^{11}}\;,
\end{eqnarray}
and 
\begin{eqnarray}\label{UUU}
U&=&1-\frac{2M}{r}+\frac{Q^2}{r^2}+\frac{\frac{2}{3}a_3Q^3}{r^4}+\frac{\frac{1}{5}Q^4\tilde{a}_4}{r^6}\nonumber\\&&+\frac{\frac{2}{7}Q^5\tilde{a}_5}{r^8}+\frac{\frac{2}{9}Q^6\tilde{a}_6}{r^{10}}+\frac{\frac{2}{11}Q^7\tilde{a}_7}{r^{12}}\;.
\end{eqnarray}
Here $\tilde{a}_i$ are the rescaled $a_i$. The spacetime can have at most seven horizons. For example, when 
\begin{eqnarray}
M&=&1\;,\ \ Q=0.9\;, \ \ a_3=-0.01\;, \ \ \tilde{a}_4=10^{-5}\;, \ \ \nonumber\\ \tilde{a}_5&=&-10^{-10}\;, \ \ \tilde{a}_6=10^{-16}\;, \ \ \tilde{a}_7=-10^{-23}\;,
\end{eqnarray}
we obtain from the horizon equation $U=0$
\begin{eqnarray}
r_i&=&-0.069\;,\ \ \ -0.0164\;,\ \ \ -0.00357\;,\ \ \ -0.0008\;,\nonumber\\&&\ \ -0.00029\;,\ \ 0.000289\;,\ \  0.000809\;,\ \ 0.00357\;,\nonumber\\&&
\ \ 0.0164\;,\ \ 0.085\;,\ \ 0.546\;,\ \ 1.44\;.
\end{eqnarray}
The seven positive roots correspond to the positions of seven horizons. So we factorize $U$ as follows
\begin{eqnarray}
U&=&\prod_{\alpha=1}^5\left(1+\frac{k_{\alpha}}{r}\right)\prod_{\alpha=1}^{7}\left(1-\frac{h_{\alpha}}{r}\right)\;,
\end{eqnarray}
with $k_i$ and $h_i$ positive constants and $h_i$ satisfy   
\begin{eqnarray}
0<h_1<h_2<h_3<h_4<h_5<h_6<h_7\;.
\end{eqnarray}
Cvetic et al \cite{cvetic:2018} show that if one regards the Christodoulou and Ruffini
formula for the total energy or enthalpy as defining the Gibbs surface, then the rules
of Gibbsian thermodynamics imply that negative temperatures arise inevitably on some inner horizons, as does the conventional form of the first law. They propose the Gibbsian temperature of horizon-i is given by \cite{zhai:2010,cvetic:2018}
\begin{eqnarray}
T_i=\frac{1}{4\pi}U^{'}|_{r=h_i}=\frac{1}{4\pi{h_i}}\prod_{\alpha=1}^5\left(1+\frac{k_{\alpha}}{h_i}\right)\prod_{\alpha\neq{i}}^{7}\left(1-\frac{h_{\alpha}}{h_i}\right)\;.
\end{eqnarray}
So we have  
\begin{eqnarray}
T_7&=&8.51\cdot{10^{-15}}\left(\frac{10^{8}\mathrm{M}_{\bigodot}}{\mathrm{M}}\right)\mathrm{K}\;, \ \ \  \ \ \nonumber\\  T_6&=&-5.70\cdot{10^{-14}}\left(\frac{10^{8}\mathrm{M}_{\bigodot}}{\mathrm{M}}\right)\mathrm{K}\;, \nonumber\\  T_5&=&3.54\cdot{10^{-11}}\left(\frac{10^{8}\mathrm{M}_{\bigodot}}{\mathrm{M}}\right)\mathrm{K}\;, \nonumber\\ 
T_4&=&-1.42\cdot{10^{-7}}\left(\frac{10^{8}\mathrm{M}_{\bigodot}}{\mathrm{M}}\right)\mathrm{K}\;, \nonumber\\  T_3&=&6.02\cdot{10^{-3}}\left(\frac{10^{8}\mathrm{M}_{\bigodot}}{\mathrm{M}}\right)\mathrm{K}\;, \nonumber\\  T_2&=&-3.54\cdot{10^{3}}\left(\frac{10^{8}\mathrm{M}_{\bigodot}}{\mathrm{M}}\right)\mathrm{K}\;, \nonumber\\
T_1&=&3.07\cdot{10^{8}}\left(\frac{10^{8}\mathrm{M}_{\bigodot}}{\mathrm{M}}\right)\mathrm{K}\;.  
\end{eqnarray}
We wee that some of the inner horizons have positive temperature but some do have negative temperature. It is well-known that absolute temperature is classically bound to be positive. However, under some special quantum conditions, negative temperatures—in which high-energy states are more occupied than low-energy states—are also possible. Such states have been realized
in localized spin systems \cite{negative:1951,negative:1997,negative:2011}. Thus the inner horizons with negative temperature would then be thought of as the analogue of
a spin system. Volovik \cite{vol:2021} thinks that every positive temperature  determines the rate of the tunneling from $r_{*}=-\infty$ to $r_{*}=+\infty$, while every negative temperature determines the occupation number of particles in the vicinity of $r_{*}=-\infty$.  Physically, the positive or negative of temperature for the horizon is closely related to the fact that an attractive force or repulsive force detected by an observer from the horizon. The deeper we go toward the center of the black hole, the higher the magnitude of temperature for the inner horizons. For black holes with many horizons, it can be inferred that the temperature of the inner-most horizon will be astonishingly high. 

In theory,  the original suggestion that inner horizons should be assigned a negative temperature \cite{curir:1979} 
was based on not quantum field theory, but quantum mechanical systems, such as spin systems, exhibiting population inversion \cite{ramsey:1956,pur:1951}.  This proposal of  negative temperature is consistent with the laws of thermodynamics \cite{swen:2015,fren:2014}. On the ohther hand, through the correlation
between temperature and surface gravity,  Tiandho {\cite{tiandho:2017}} concluded that the temperature of
the inner horizon is always negative and that of the outer horizon is always
positive for Reisnner-Nordstrom black holes.  Wu \cite{wu:2004} alos proposed that the
inner horizon’s temperature is negative because of its negative surface gravity. 
Finally and experimentally, by tailoring the Bose-Hubbard Hamiltonian,  S. Braun et al \cite{braun:2013} created an attractively interacting ensemble of ultracold bosons at negative temperature.

Now we turn to the Bekenstein-Hawking entropy which is associated to horizon-i as follows
\begin{eqnarray}
S_i={\pi}h_i^2\;.
\end{eqnarray}
The corresponding Coulomb potential on horizon-i is 
\begin{eqnarray}
\Phi_i=\phi|_{r=h_i}\;.
\end{eqnarray}
In order to derive the first law of thermodynamic, we focus on the equation of horizons. 
\begin{eqnarray}
U|_{r=h_i}=0\;,
\end{eqnarray}
from which we obtain
\begin{eqnarray}
M&=&\frac{1}{2}h_i+\frac{Q^2}{2h_i}+\frac{\frac{1}{3}a_3Q^3}{h_i^3}+\frac{\frac{1}{10}Q^4\tilde{a}_4}{h_i^5}+\frac{\frac{1}{7}Q^5\tilde{a}_5}{h_i^7}
\nonumber\\&&+\frac{\frac{1}{9}Q^6\tilde{a}_6}{h_i^{9}}+\frac{\frac{1}{11}Q^7\tilde{a}_7}{h_i^{11}}\;,
\end{eqnarray}
or 
\begin{eqnarray}
M&=&\frac{1}{2}\left(\frac{S_i}{\pi}\right)^{\frac{1}{2}}+\frac{Q^2}{2}\left(\frac{\pi}{S_i}\right)^{\frac{1}{2}}+\frac{a_3Q^3}{3}\left(\frac{\pi}{S_i}\right)^{\frac{3}{2}}\nonumber\\&&+\frac{Q^4\tilde{a}_4}{10}\left(\frac{\pi}{S_i}\right)^{\frac{5}{2}}+\frac{Q^5\tilde{a}_5}{7}\left(\frac{\pi}{S_i}\right)^{\frac{7}{2}}+\frac{Q^6\tilde{a}_6}{9}\left(\frac{\pi}{S_i}\right)^{\frac{9}{2}}\nonumber\\&&+\frac{Q^7\tilde{a}_7}{11}\left(\frac{\pi}{S_i}\right)^{\frac{11}{2}}\;.
\end{eqnarray}
Now the energy $M$ is regarded as a function of the extensive variables, $S_i$, $Q$, $a_3$, $\tilde{a}_4$, $\tilde{a}_5$, $\tilde{a}_6$, $\tilde{a}_7$. Differentiating above relation, we find the first-law of thermodynamics  
\begin{eqnarray}
dM&=&\frac{\partial{M}}{\partial{S_i}}dS_i+\frac{\partial{M}}{\partial{Q}}dQ+\frac{\partial{M}}{\partial{a_3}}da_3+\frac{\partial{M}}{\partial{\tilde{a}_4}}d\tilde{a}_4\nonumber\\&&+\frac{\partial{M}}{\partial{\tilde{a}_5}}d\tilde{a}_5+\frac{\partial{M}}{\partial{\tilde{a}_6}}d\tilde{a}_6+\frac{\partial{M}}{\partial{\tilde{a}_7}}d\tilde{a}_7\;\nonumber\\&=&T_idS_i+\Phi_idQ+J_3da_3+J_4d\tilde{a}_4+J_5d\tilde{a}_5\nonumber\\&&+J_6d\tilde{a}_6+J_7d\tilde{a}_7\;,
\end{eqnarray}
where 
\begin{eqnarray}
J_3&=&\frac{Q^3}{3h_i^3}\;,\ \ J_4=\frac{Q^4}{10h_i^5}\;,\ \ J_5=\frac{Q^5}{7h_i^7}\;,\nonumber\\ \ \ J_6&=&\frac{Q^6}{9h_i^9}\;,\ \ J_7=\frac{Q^7}{11h_i^{11}}\;, 
\end{eqnarray}
and $T_i$, $\Phi_i$ are exactly that defined previously. 
It is found that the Smarr relation 
\begin{eqnarray}
M&=&2T_iS_i+\Phi_iQ+J_3a_3+2J_4\tilde{a}_4+3J_5\tilde{a}_5\nonumber\\&&+4J_6\tilde{a}_6+5J_7\tilde{a}_7\;, 
\end{eqnarray}
holds on every single horizon. 

When the cosmological constant is  taken into account,  the first law is generalized to 
\begin{eqnarray}
dM&=&T_idS_i+\Phi_idQ+J_3da_3+J_4d\tilde{a}_4+J_5d\tilde{a}_5\nonumber\\&&+J_6d\tilde{a}_6+J_7d\tilde{a}_7+\mathbb{V} d\mathbb{P}\;,
\end{eqnarray}
where the pressure and the thermodynamic volume are defined as usual \cite{kastor:2009, cvetic:2011} 
\begin{eqnarray}
\mathbb{P}=-\frac{\Lambda}{8\pi}\;,\ \ \ \ \mathbb{V}=\frac{4\pi{r_i^2}}{3}\;,
\end{eqnarray}
The Smarr relation turns out to be
\begin{eqnarray}
M&=&2T_iS_i+\Phi_iQ+J_3a_3+2J_4\tilde{a}_4+3J_5\tilde{a}_5\nonumber\\&&+4J_6\tilde{a}_6+5J_7\tilde{a}_7-2\mathbb{V} \mathbb{P}\;.
\end{eqnarray}

\section{motion of test particles}
In this section, we  consider the motion of radially moving massive particles in the background of multi-horizon black hole spacetime. To simplify our discussion, we will assume that the particles are electrically neutral and with unit  rest mass. In this case, the particles will follow geodesics. For electrically charged particles, one must take into account the Lorenz force on the particles produced by the black hole. 

\subsection{classical motion}

\begin{figure}
	\includegraphics[width=7cm,height=5cm]{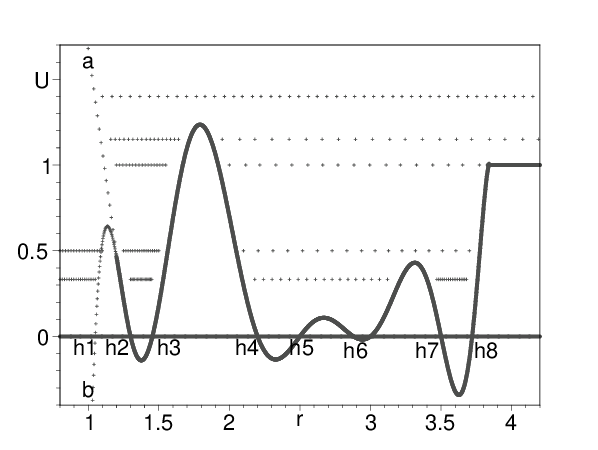}
	\caption{A schech of effective potential $U$ with zero angular momentum.} 
	\label{eff.eps}
\end{figure}

In this subsection, we consider the classical motion of test particles. The equation of motion for neutral particle is then given by 
\begin{equation}\label{classical}
\dot{r}^2+U\left(1+\frac{L^2}{r^2}\right)=E^2\;,
\end{equation}
where $E$ and $L$ are the energy and angular momentum of the particle, respectively. This clearly shows the sum of kinetic energy and potential energy is a constant  where 
\begin{equation}
V_{eff}=U\left(1+\frac{L^2}{r^2}\right)\;,
\end{equation}
plays the role of an effective  potential. Qualitative analysis on the radial motion of particles can be obtained directly by simply plotting the potential $U$ as shown in Fig.~(\ref{eff.eps}).  Qualitative analysis indicates that the motion with angular momentum motion is similar to the radial motion. Therefore, it is enough to merely give an analysis on radial motion.  As an example, we consider 7-horizon and 8-horizon black holes, respectively,  i.e. $h_1<h_2<h_3<h_4<h_5<h_6<h_7<h_8$.  For 7-horizon black holes, the potential approaches positive infinity in the vicinity of singularity as shown by dotted line $a$. In contrast, the potential approaches negative infinity near the singularity as shown by dotted line $b$. We see there exists radial boundaries for the motion of the particle depending on the value of energy squared  $E^2$,  as indicated with pointed lines. We shall make a discussion in three situations, i.e. $E^2=1$, $E^2>1$ and $E^2<1$, respectively. In the first place, the case $E^2=1$ corresponds to the particle being released from rest at infinity.  For a 7-horizon black hole, the particle with energy $E^2=1$ would cross $h_8,h_7,h_6,h_5,h_4$ successively. Then it encounters a potential barrier and rebounds. Subsequently, the particle crosses the horizons $h_4,h_5,h_6,h_7,h_8$ in chronological order. In the end, it escapes to infinity again. We emphasize that this process is only experienced by the free-falling observer. The distant observer far away from the black hole will never observe this process.  On the other hand, if the particle is inside of the black hole, it would oscillate in some region. This physical process is also only undergone by the particle itself and the distant observer is unable to detect the process.  Secondly, when $E^2>1$ and for 7-horizon black hole, the particle always cannot reach the central singularity  but is instead repelled, passing back through all the horizons and ultimately emerges in a another asymptotically flat spacetime which is different from ours. For sufficient large energy $E^2>1$ and for 8-horizon black hole, the particle eventually falls into the singularity. Finally, we consider the case of $E^2<1$. In this case, for both 7-horizon and 8-horizon black holes, there always exists local minimum and local maximum of the radial potion for particles. This indicates that those neutral particles always  oscillate between their maximum and minimum distance. In this situation, many particles are bounded in the interior of black holes which is very similar to that electrons are bounded in an atom. We emphasize once again that due to the presence of event horizon, all the precesses happened in the interior of black hole will never be seen by the distant observer. However, it is not the case from the perspective of quantum mechanics.

\vspace{-8pt}

\subsection{quantum motion}
In this subsection, we consider the quantum motion of test particles. Eq.~(\ref{classical})  describes a one-dimensional classical motion of test particles. By contrast, the equation of motion for the quantum motion is given by the stationary Schrodinger equation   
\begin{equation}\label{quantum}
-\frac{1}{2}\frac{d^2\Psi}{dr^2}+V_{eff}\Psi=\omega\Psi\;,
\end{equation}
where $\omega$ is the energy of the particle. We note that here the energy of quantum particle can be negative, i.e $\omega<0$. However, the energy of classical particle discussed in the  above subsection is always positive, i.e. $E>0$.   Because there are many potential wells, the bound states for particle exist and the discrete energy of particles can be  given by the WKB method, 

\begin{equation}\label{enenrgy}
\left(n+\frac{1}{2}\right)\pi=\int_{a}^{b}\sqrt{2\left(\omega-V_{eff}\right)}dr\;.
\end{equation}
Here $a$ and $b$ stand for the transition points.  On the other hand, there are also many potential barriers. Using the WKB method, the transmission probability of particle is given by 
\begin{equation}\label{enenrgy}
\mathbb{T}=e^{-2|\int_{a}^{b}\sqrt{2\left(\omega-V_{eff}\right)}dr|}\;.
\end{equation}
Therefore, from the view of point of quantum mechanics, a particle inside of the black hole can escape from the black hole to the spacial infinity.

\section{the motion of Klein-Gordon field}
In this section, we  consider the motion of Klein-Gordon field in the background of multi-horizon black hole spacetime. The field is regarded as a test scalar field. This means the field makes no contribution to the curvature of spacetime manifold. By studying the motion of test scalar field, we can obtain the informations of curved spacetime background just like we can solve for the solar mass from the motion of Earth. The equation of motion of Klein-Gordon equation is

\begin{eqnarray}\label{KG}
\nabla_{\mu}\nabla^{\mu}\varphi-\mu_0^2\varphi=0\;.
\end{eqnarray}
where $\mu_0$ is the mass of test particles. Separating $\varphi$ as follows
\begin{eqnarray}
\varphi=\frac{e^{-i\omega{t}}\psi\left(r\right)}{r}Y_{lm}\;,
\end{eqnarray}
with $Y_{lm}$ the spherical harmonic function. Defining the tortoise coordinate 
\begin{eqnarray}
r_{*}&=&\int{\frac{dr}{U}}\nonumber\\
&=&r-\sum_{\beta=1}^5\frac{k_{\beta}^{12}\ln|r+k_{\beta}|}{\prod_{\alpha\neq{\beta}}^5\left(k_{\beta}-k_{\alpha}\right)\prod_{\alpha=1}^7\left(k_{\beta}+h_{\alpha}\right)}\nonumber\\&&
+\sum_{\beta=1}^7\frac{h_{\beta}^{12}\ln|r-h_{\beta}|}{\prod_{\alpha=1}^5\left(h_{\beta}+k_{\alpha}\right)\prod_{\alpha\neq{\beta}}^7\left(h_{\beta}-h_{\alpha}\right)}\;.
\end{eqnarray}

\begin{figure}
	\includegraphics[width=7cm,height=5cm]{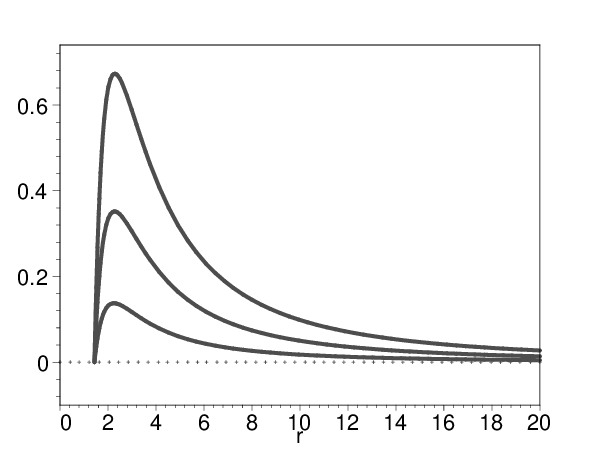}
	\caption{There  are potential barriers (on the order of $10^{0}$) in $[h_7\;, +\infty)$. $l=3,2,1$ are put from top to down, respectively. } 
	\label{pot-7.eps}
\end{figure}

\begin{figure}
	\includegraphics[width=7cm,height=5cm]{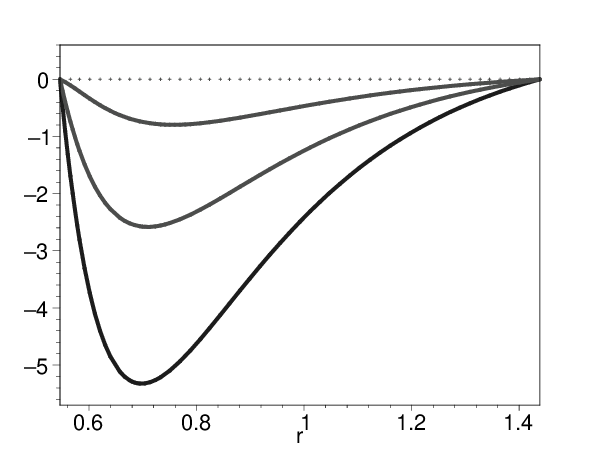}
	\caption{There are potential wells  (on the order of $10^{0}$) in $[h_6\;, h_7]$. $l=1,2,3$ are put from top to down, respectively.  } 
	\label{pot-67.eps}
\end{figure}

\begin{figure}
	\includegraphics[width=7cm,height=5cm]{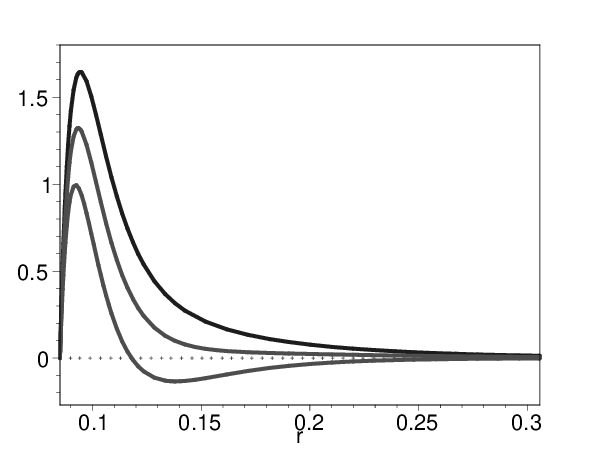}
	\caption{Potential barriers coexist  with potential wells (on the order of $10^{5}$) in $[h_5\;, h_6]$. $l=7,5,1$ are set from top to down, respectively.  } 
	\label{pot-56.eps}
\end{figure}

\begin{figure}
	\includegraphics[width=7cm,height=5cm]{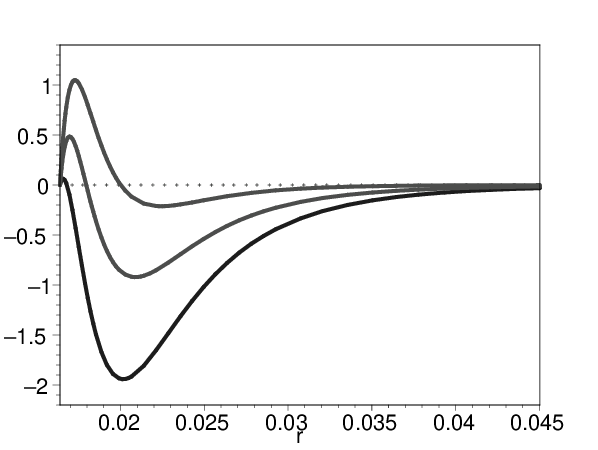}
	\caption{Potential barriers coexist  with potential wels (on the order of $10^{12}$) in $[h_4\;, h_5]$. $l=1,200,300$ from top to down.  } 
	\label{pot-45.eps}
\end{figure}

\begin{figure}
	\includegraphics[width=7cm,height=5cm]{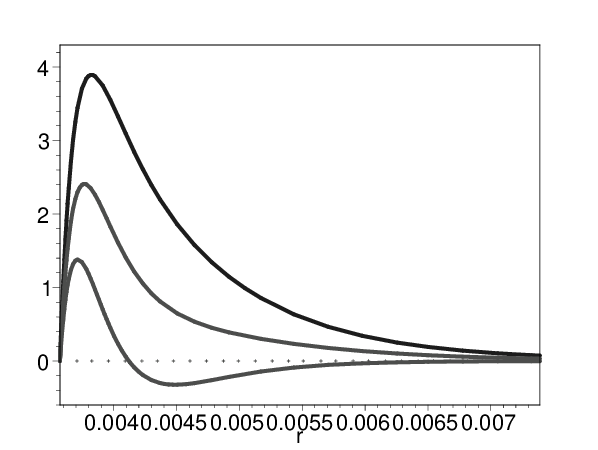}
	\caption{Potential barriers coexist  with potential wells (on the order of $10^{21}$) in $[h_3\;, h_4]$. $l=30000,20000,1$ from top to down.  } 
	\label{pot-34.eps}
\end{figure}

\begin{figure}
	\includegraphics[width=7cm,height=5cm]{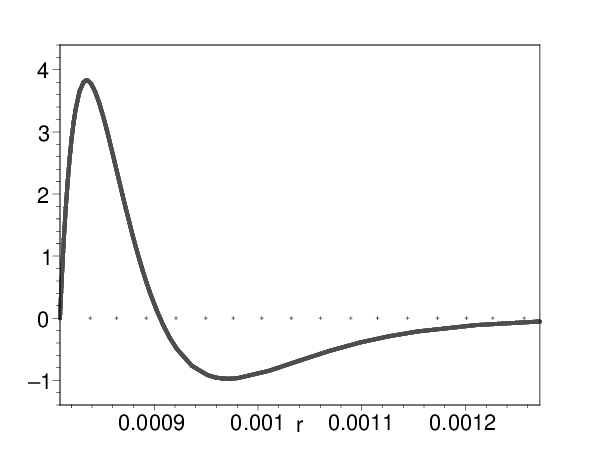}
	\caption{Potential barriers coexist with potential wells (on the order of $10^{32}$) in $[h_2\;, h_3]$. We set  $l=1,1000,10000$.  It shows the effect of $l$ on the potential is negligible. } 
	\label{pot-23.eps}
\end{figure}

\begin{figure}
	\includegraphics[width=7cm,height=5cm]{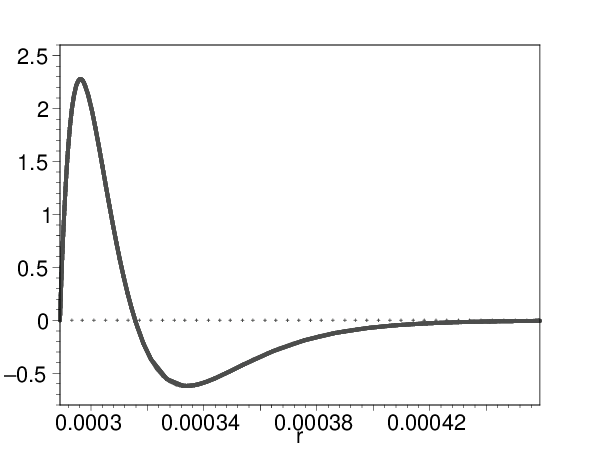}
	\caption{Potential barriers coexist  with potential wells (on the order of $10^{42}$) in $[h_1\;, h_2]$. We put $l=1,1000,10000$.  It shows the effect of $l$ on the potential is negligible.}
	\label{pot-12.eps}
\end{figure}

\begin{figure}
	\includegraphics[width=7cm,height=5cm]{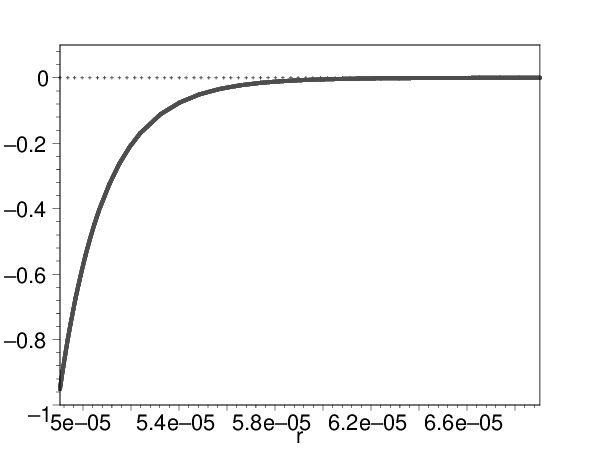}
	\caption{There are infinite deep potential wells  (on the order of $10^{65}$) in $[0\;, h_1]$. We put $l=1,1000,10000$. It shows the effect of $l$ on the potential is negligible. } 
	\label{pot-01.eps}
\end{figure}

We conclude that
\begin{eqnarray}
&&r=+\infty\ \ \ \Leftrightarrow\ \ {r_{*}=+\infty}\;, \nonumber\\&& r=h_7\ \ \ \ \Leftrightarrow\ \ {r_{*}=-\infty}\;,\ \ \ \nonumber\\&&
r=h_6\ \ \ \ \ \Leftrightarrow\ \ {r_{*}=+\infty}\;,\ \nonumber\\&& r=h_5\ \ \ \ \ \Leftrightarrow\ \ {r_{*}=-\infty}\;,\ \ \  \nonumber\\&&
r=h_4\ \ \ \ \ \Leftrightarrow\ \ {r_{*}=+\infty}\;
,\nonumber\\&&  r=h_3\ \ \ \ \ \Leftrightarrow\ \ {r_{*}=-\infty}\;,\ \ \  \nonumber\\&&
r=h_2\ \ \ \ \ \Leftrightarrow\ \ {r_{*}=+\infty}\;,\nonumber\\&& r=h_1\ \ \ \ \ \Leftrightarrow\ \ {r_{*}=-\infty}\;,
\nonumber\\&& r=0\ \ \ \ \ \ \ \Leftrightarrow\ \ {r_{*}=0}\;.
\end{eqnarray}
Therefore, the space is divided into eight regions, i.e. 
\begin{eqnarray}
&&(1)\ \ (+\infty\;,h_7]\;, \ \ \ \ \ \  (2)\ \ [h_7\;, h_6]\;,\ \ \ \nonumber\\&&
(3)\ \ [h_6\;, h_5]\;, \ \ \ \ \ \ \ \ \  (4)\ \ [h_5\;, h_4]\;,\ \ \ \nonumber\\&&
(5)\ \ [h_4\;, h_3]\;, \ \ \ \ \ \ \ \ \  (6)\ \ [h_3\;, h_2]\;,\ \ \ \nonumber\\&&
(7)\ \ [h_2\;, h_1]\;, \ \ \ \ \ \ \ \ \  (8)\ \ [h_1\;, 0)\;.
\end{eqnarray}
In the tortoise coordinate system, the equation of motion takes the form   
\begin{eqnarray}\label{kangzoubukangzou}
\frac{\partial^2\psi}{\partial{r}_{*}^2}=\left(V-\omega^2\right)\psi\;,
\end{eqnarray}
where prime denotes the derivative with respect to $r$. The potential $V$ is defined as  
\begin{eqnarray}
V=U\left[\frac{U^{'}}{r}+\frac{l\left(l+1\right)}{r^2}+\mu_0^2\right]\;.
\end{eqnarray}
This equation is similar in form to the stationary Schrodinger equation after we regard $r$ as the function of $r_{*}$. Without the loss of generality, we consider the massless case $\mu_0=0$.  Then the potential is vanishing in both spatial infinity and all the horizons. Thus we can solve  the equation of motion in every region with the boundary conditions   
\begin{eqnarray}
\psi \sim e^{\pm{i}\omega{r}_{*}}\;,
\end{eqnarray}
at spatial infinity and all the horizons. Here $+$ and $-$ denotes the purely outgoing and ingoing wave, respectively.  From Fig.~\ref{pot-7.eps} to Fig.~\ref{pot-01.eps}, we plot the potential for every region. 

In region (1) or $[h_7\;, +\infty)$, the potential is always positive and behaves as a barrier which climbs even higher with the increasing of angular quantum number $l$.  There are four important processes in this region. In the first place, quantum perturbations in this region is stable because the potential is positive everywhere.  This is shown by Vincent Moncrief  in the case of Reissner-Nordstrome black hole \cite{RN:1974}. Secondly,  reflection and transmission phenomena will occur in this region. Thirdly, Hawking radiation is produced in this region.   Finally,  the quasinormal modes are  also produced here.  

In region (2) or $[h_6\;, h_7]$, the potential behaves as a well which gets even deeper with the increasing of angular quantum number $l$.  In this region, three important processes will occur. The first one  is the transition and  reflection and the second  is the existence of bound states of waves. The third  is the trapping of  fields .  The  trapping process can be outlined as follows. The equation of motion Eq.~(\ref{kangzoubukangzou}) tells us if the energy $\omega$ is purely imaginary 
\begin{eqnarray}
\omega={i}\omega_0\;,
\end{eqnarray}
with $\omega_0$ positive constant and    
\begin{eqnarray}
V-\omega^2=V+\omega_0^2>0\;,
\end{eqnarray}
the fields decay exponentially in $r_{*}$ as $\left|{r_{*}}\right|=+\infty$ with the boundary conditions
\begin{eqnarray}
\psi \sim e^{-{i}\omega{r}_{*}}=e^{\omega_0r_{*}}\sim{0}\;,  \ \ \ \ at    \ \ \ r_{*}=-\infty\;, \nonumber\\
\psi \sim e^{+{i}\omega{r}_{*}}=e^{-\omega_0r_{*}}\sim{0}\;,  \ \ \ \ at    \ \ \ r_{*}=+\infty\;.
\end{eqnarray}
 This means the fields are trapped in the potential well.  On the other hand, the fields decrease exponentially (the increasing mode is unphysical and thus given up) in time which means the they are absorbed by the potential well.  With the trapping of fields in the well, this region becomes increasing unstable.  Eventually, the potential well is filled and leveled up with energies.  Then the inner horizon disappears.

In region (3), the potential behaves as a barrier for large $l$. For small $l$, a potential well gradually appears and coexists with the barrier. The emergence of  negative gap in the potential leads to eikonal instability according to \cite{tak:2009}. We expect the sinking and inflation of waves would happen in the well.  We note that the height of the potential is $5$ orders of magnitude  of region (1) and region (2).  Similar cases occur for regions (4,5,6,7). Region (8) gives us an infinite deep potential well. It is remarkable that the effect of $l$ is negligible in the regions (6,7,8) .

\vspace{80pt}

Now let's consider a wave incident on the black hole from spatial infinity.  Firstly, the wave encounters a potential barrier with the height on the order of $10^{0}$ near the event horizon. It undergoes reflection and transmission. The wave that passes through $h_7$ then encounters a potential well with the depth on the order of $10^{0}$ and undergoes reflection and transmission again. In this region, the capture of  waves would happen. The surviving wave then passes through $h_6$ and encounters a high potential barrier (on the order of $10^{5}$) , continues to undergo reflection and transmission. For small angular quantum numbers, the wave will first encounter a potential well, followed by a potential barrier. After passing through the horizon $h_5$, the wave will encounter a deeper potential well, and after passing through the horizon $h_4$, the wave will encounter a higher potential barrier again. The further inside, the less significant the effect of angular quantum numbers becomes. In fact, the influence of large or small angular quantum numbers on the potential is no longer apparent. Finally, we see that the potential barrier becomes  extremely  high and the potential well becomes extremely deep. As the subsequent potential barriers become higher and higher,  it will become increasingly difficult for the wave to reach the center of the black hole. 
\section{avoiding the mass inflation problem}
In the above sections,  we have derived and studied the physics of multi-horizon black holes.  We find that  many interesting things would happen for the multi-horizon black holes, such as the alternate occurrence of positive and negative temperatures, the alternate occurrence of vast potential barriers and potential wells and the  various interesting motions for classical particles.  Although the physics is interesting, but the solutions suffer from  the mass inflation problem. So in the last section, we shall improve the multiple black hole solutions to that without mass inflation by adjusting the inner horizons to overlap.  We summarize them in the below.  
\subsection{Three-horizon black holes as the mother solution }
\textbf{1. One-horizon black holes}

The metric function of one-horizon black holes is given by
\begin{eqnarray}
U=\left(1+\frac{M}{4r}\right)\left(1-\frac{3M}{4r}\right)^3\;,
\end{eqnarray}
with $M$ the mass of black hole.  The solution comes from the three-horizon black holes with the three horizons coincide, Eq.~(\ref{eq-oh}).  In this case, both the surface gravity and temperature of the single horizon is vanishing. The singularity is spacelike.

\textbf{2. Two-horizon black holes }

The metric function of two-horizon black holes is given by
\begin{eqnarray}
&&U=\left[1+{\frac{s\tilde{h}_1}{\left(1+2s\right)r}}\right]\left(1-\frac{\tilde{h}_1}{r}\right)^2\left(1-\frac{s\tilde{h}_1}{r}\right)\;,
\end{eqnarray}
with $\tilde{h}_1>0$ and $s>1$. It is also coming from the three-horizon black holes, Eq.~(\ref{eq-th}). In this case, we have two horizons. One of them is the outer event horizon and the other is the inner  coincident horizon. Both the surface gravity and temperature of the inner horizon are zeros. While the outer event horizon has the positive surface gravity and temperature.  The singularity is spacelike. 

\subsection{Four-horizon black holes as the the mother solution }

\textbf{1. Two-horizon black holes}

 Now we study the metric function of four-horizon black holes 
\begin{eqnarray}
U=1-\frac{2M}{r}+\frac{Q^2}{r^2}+\frac{\bar{a}_3}{r^4}+\frac{\bar{a}_4}{r^6}\;.
\end{eqnarray}
It has four parameters, $M, Q,\bar{a}_3,\bar{a}_4$. We can  factorize it  in  the form  
\begin{eqnarray}
U=\left(1+\frac{\bar{k}_1}{r}+\frac{\bar{k}_2}{r^2}\right)\left(1-\frac{\bar{h}}{r}\right)^3\left(1-\frac{s\bar{h}}{r}\right)\;.
\end{eqnarray}

Then we conclude that 
\begin{eqnarray}
\bar{k}_1&=&\frac{\left(1+3s\right)^2\bar{h}}{3+8s^2+9s}\;,  \ \ \ \  \ \ \bar{a}_3=-{\frac{\bar{h}^4\left(3s+10s^3+6s^2+1\right)}{3+8s^2+9s}}\;,\ \ \  
\nonumber\\M&=&\frac{4\bar{h}_1\left(1+s\right)^3}{3+8s^2+9s}\;,\ \ \ \ \  \  Q^2=\frac{3\bar{h}^2\left(2+6s+7s^2+5s^3\right)}{3+8s^2+9s}\;,\ \ \  
\nonumber\\\bar{k}_2&=&\frac{\left(1+3s\right)\bar{h}^2s}{3+8s^2+9s}\;,\ \ \ \ \ \bar{a}_4=\frac{s^2\bar{h}^6\left(1+3s\right)}{3+8s^2+9s}\;.
\end{eqnarray}
Now the four-parameter black holes ($M, Q,\bar{a}_3,\bar{a}_4$) becomes two-parameter one ($s,\bar{h}$). When 
\begin{eqnarray}
\bar{h}>0\;,\ \ \ \  s>1\;, 
\end{eqnarray}
the inner three horizons coincide with the vanishing surface gravity. In this case, we are left with two horizons and the singularity is timelike. On the other hand, if
\begin{eqnarray}
\bar{h}>0\;,\ \ \ \  0<s<1\;, 
\end{eqnarray}
the outer three horizons coincide with he vanishing surface gravity. In this case,  the surface gravity (and the temperature) is positive for the surviving inner horizon. Therefore, the mass inflation phenomenon remains.   

\textbf{2. One-horizon black holes}

When 
\begin{eqnarray}
\bar{h}>0\;,\ \ \ \  s=1\;, 
\end{eqnarray}
the inner three horizons and the outer event horizon coincide also with the vanishing of surface gravity. In this case, the singularity is timelike. 

\textbf{3. Two-horizon black holes}

Finally, we factorize the metric function as follows

\begin{eqnarray}
U=\left(1+\frac{\bar{k}_1}{r}+\frac{\bar{k}_2}{r^2}\right)\left(1-\frac{\bar{h}_1}{r}\right)^2\left(1-\frac{\bar{h}_2}{r}\right)^2\;,
\end{eqnarray}
and conclude that 
\begin{eqnarray}
\bar{k}_1&=&\frac{2\left(\bar{h}_1+\bar{h}_2\right)\bar{h}_2\bar{h}_1}{\bar{h}_2^2+3\bar{h}_2\bar{h}_1+\bar{h}_1^2}\;, \ \ \ \ \  
\bar{k}_2=\frac{\bar{h}_2^2\bar{h}_1^2}{\bar{h}_2^2+3\bar{h}_2\bar{h}_1+\bar{h}_1^2}\;,\ \ \  
\nonumber\\M&=&\frac{\left(\bar{h}_1+\bar{h}_2\right)^3}{\bar{h}_2^2+3\bar{h}_2\bar{h}_1+\bar{h}_1^2}\;,\ \ \ \ \ \bar{a}_4=\frac{\bar{h}_2^4\bar{h}_1^4}{\bar{h}_2^2+3\bar{h}_2\bar{h}_1+\bar{h}_1^2}\;,\nonumber\\  Q^2&=&\frac{\bar{h}_2^4+\bar{h}_1^4+3\bar{h}_2^3\bar{h}_1+7\bar{h}_1^2\bar{h}_2^2+3\bar{h}_1^3\bar{h}_2}{\bar{h}_2^2+3\bar{h}_2\bar{h}_1+\bar{h}_1^2}\;,\ \nonumber\\\bar{a}_3&=&-\frac{\bar{h}_1^2\bar{h}_2^2\left(\bar{h}_1\bar{h}_2+2\bar{h}_2^2+2\bar{h}_1^2\right)}{\bar{h}_2^2+3\bar{h}_2\bar{h}_1+\bar{h}_1^2}\;,
\end{eqnarray}
with two parameters

\begin{eqnarray}
\bar{h}_1>0\;,\ \ \ \ \bar{h}_2>0\;.
\end{eqnarray}

In this situation, we are left with  two horizons both with the vanishing surface gravity. The singularity here is timelike. 

Similarly, taking the five-horizon black holes, six-horizon black holes and so on as the mother solutions, one can produce many other black holes without mass inflation.   

\section{conclusion and discussion}
The well-known Reissner-Nordstr$\ddot{o}$m black holes,  Kerr black holes and Kerr-Newman black holes are all suffering from the mass inflation problem.  Surface cause of the problem is that they have two horizons. Underlying cause is that the observer at rest  on the event horizon feels an infinite attractive force while  the observer at rest  on the inner horizon feels an infinite repulsive force.  This is also the reason why the surface gravity of outer event horizon is positive while the surface gravity is negative for inner horizon.  For free falling particles, the inner horizon is an insurmountable high wall. Therefore, particles falling into a black hole and the resulting energy density will accumulate in the vicinity of the inner horizon.  Thus the phenomenon of mass-inflation occurs. Noticing that  the mass inflation exponentially  grows with the power of  absolute value of surface gravity, Carballo-Rubio et al \cite{rubio:2022} showed that, if we are able to let the surface gravity of inner horizons vanishes,  then the exponential growth of mass inflation instability would not happen.  It is indeed the case. In fact, the extreme black holes with  two horizons coincident have the vanishing surface gravities such that the mass inflation does not exist.  On the other hand, Cai \cite{cai:1999} investigated the stability of the  coincident Cauchy horizon. It is found that, despite the asymptotic behavior of spacetimes (flat, anti-de Sitter, or de Sitter),  this kind of  Cauchy horizon is stable against the classical perturbations. 
In view of this point, we start our exploration of multiple-horizon black holes. 

In the framework of Einstein nonlinear-Maxwell field theories,  we obtain the black hole solutions in the most general form (Eq.~(\ref{U})).  In the solution, two integration constants are present and they play the role  of physical mass and electric charge, respectively. The other parameters $a_i$ come from the theories. By adjusting the theoretical parameters $a_i$, the infinite series in the metric function becomes finite one and various multiple-horizon black holes are found.  

In the process of study for the thermodynamics of multiple horizon black holes, we find that the positive and negative surface gravities or temperatures appear alternately. It is exactly this character makes us to be able to achieve black holes without mass inflation.  According to the proposal of  Cvetic et al \cite{cvetic:2018} , these negative temperatures should be named after Gibbsian temperatures. 

By studying the motion of test particles and test fields,  we find a phenomenon of alternate occurrence of
the vast number of  potential barriers and potential wells. Therefore, imposed different boundary conditions, many quantum mechanic phenomena such as the reflection, transition, quantum stable states and so on are expected to occur. Except this,  various interesting motions for classical particles appear.  In short, the physics of multiple horizon black holes is considerable profuse.  Finally, as another strategy to avoid mass-inflation, Hale et al. \cite{hale:2025} eliminate the Cauchy horizon by regularizing the self energy of point charges and Casadio et al. \cite{cas:2015,cas:2017} used the method of horizon quantum mechanics to demonstrate that the probability of achieving an inner horizon is lower than the probability of achieving an outer event horizon.

\section*{ACKNOWLEDGMENTS}

The work is partially supported by the Special Exchange Program of CAS, National Key RD Program of China grants No.
2022YFF0503404, No. 2022SKA0110100 and the Central Guidance for Local Science and Technology Development Fund Project with Grand No. 2024ZY0113.

\newcommand\arctanh[3]{~arctanh.{\bf ~#1}, #2~ (#3)}
\newcommand\ARNPS[3]{~Ann. Rev. Nucl. Part. Sci.{\bf ~#1}, #2~ (#3)}
\newcommand\AL[3]{~Astron. Lett.{\bf ~#1}, #2~ (#3)}
\newcommand\AP[3]{~Astropart. Phys.{\bf ~#1}, #2~ (#3)}
\newcommand\AJ[3]{~Astron. J.{\bf ~#1}, #2~(#3)}
\newcommand\GC[3]{~Grav. Cosmol.{\bf ~#1}, #2~(#3)}
\newcommand\APJ[3]{~Astrophys. J.{\bf ~#1}, #2~ (#3)}
\newcommand\APJL[3]{~Astrophys. J. Lett. {\bf ~#1}, L#2~(#3)}
\newcommand\APJS[3]{~Astrophys. J. Suppl. Ser.{\bf ~#1}, #2~(#3)}
\newcommand\JHEP[3]{~JHEP.{\bf ~#1}, #2~(#3)}
\newcommand\JMP[3]{~J. Math. Phys. {\bf ~#1}, #2~(#3)}
\newcommand\JCAP[3]{~JCAP {\bf ~#1}, #2~ (#3)}
\newcommand\LRR[3]{~Living Rev. Relativity. {\bf ~#1}, #2~ (#3)}
\newcommand\MNRAS[3]{~Mon. Not. R. Astron. Soc.{\bf ~#1}, #2~(#3)}
\newcommand\MNRASL[3]{~Mon. Not. R. Astron. Soc.{\bf ~#1}, L#2~(#3)}
\newcommand\NPB[3]{~Nucl. Phys. B{\bf ~#1}, #2~(#3)}
\newcommand\CMP[3]{~Comm. Math. Phys.{\bf ~#1}, #2~(#3)}
\newcommand\CQG[3]{~Class. Quantum Grav.{\bf ~#1}, #2~(#3)}
\newcommand\PLB[3]{~Phys. Lett. B{\bf ~#1}, #2~(#3)}
\newcommand\PRL[3]{~Phys. Rev. Lett.{\bf ~#1}, #2~(#3)}
\newcommand\PR[3]{~Phys. Rep.{\bf ~#1}, #2~(#3)}
\newcommand\PRd[3]{~Phys. Rev.{\bf ~#1}, #2~(#3)}
\newcommand\PRD[3]{~Phys. Rev. D{\bf ~#1}, #2~(#3)}
\newcommand\RMP[3]{~Rev. Mod. Phys.{\bf ~#1}, #2~(#3)}
\newcommand\SJNP[3]{~Sov. J. Nucl. Phys.{\bf ~#1}, #2~(#3)}
\newcommand\ZPC[3]{~Z. Phys. C{\bf ~#1}, #2~(#3)}
\newcommand\IJGMP[3]{~Int. J. Geom. Meth. Mod. Phys.{\bf ~#1}, #2~(#3)}
\newcommand\IJTP[3]{~Int. J. Theo. Phys.{\bf ~#1}, #2~(#3)}
\newcommand\IJMPD[3]{~Int. J. Mod. Phys. D{\bf ~#1}, #2~(#3)}
\newcommand\IJMPA[3]{~Int. J. Mod. Phys. A{\bf ~#1}, #2~(#3)}
\newcommand\GRG[3]{~Gen. Rel. Grav.{\bf ~#1}, #2~(#3)}
\newcommand\EPJC[3]{~Eur. Phys. J. C{\bf ~#1}, #2~(#3)}
\newcommand\EPL[3]{~Europhysics Letters.{\bf ~#1}, #2~(#3)}
\newcommand\PRSLA[3]{~Proc. Roy. Soc. Lond. A {\bf ~#1}, #2~(#3)}
\newcommand\AHEP[3]{~Adv. High Energy Phys.{\bf ~#1}, #2~(#3)}
\newcommand\Pramana[3]{~Pramana.{\bf ~#1}, #2~(#3)}
\newcommand\PTEP[3]{~PTEP.{\bf ~#1}, #2~(#3)}
\newcommand\PTP[3]{~Prog. Theor. Phys{\bf ~#1}, #2~(#3)}
\newcommand\APPS[3]{~Acta Phys. Polon. Supp.{\bf ~#1}, #2~(#3)}
\newcommand\ANP[3]{~Annals Phys.{\bf ~#1}, #2~(#3)}
\newcommand\RPP[3]{~Rept. Prog. Phys. {\bf ~#1}, #2~(#3)}
\newcommand\ZP[3]{~Z. Phys. {\bf ~#1}, #2~(#3)}
\newcommand\NCBS[3]{~Nuovo Cimento B Serie {\bf ~#1}, #2~(#3)}
\newcommand\AAP[3]{~Astron. Astrophys.{\bf ~#1}, #2~(#3)}
\newcommand\MPLA[3]{~Mod. Phys. Lett. A.{\bf ~#1}, #2~(#3)}
\newcommand\NT[3]{~Nature.{\bf ~#1}, #2~(#3)}
\newcommand\PT[3]{~Phys. Today. {\bf ~#1}, #2~ (#3)}
\newcommand\APPB[3]{~Acta Phys. Polon. B{\bf ~#1}, #2~(#3)}
\newcommand\NP[3]{~Nucl. Phys. {\bf ~#1}, #2~ (#3)}
\newcommand\JETP[3]{~JETP Lett. {\bf ~#1}, #2~(#3)}
\newcommand\PDU[3]{~Phys. Dark. Univ. {\bf ~#1}, #2~(#3)}

\end{document}